\documentclass{jjap3}
\usepackage{newtxtext,newtxmath}
\usepackage{units}
\usepackage{siunitx}
\usepackage{float}

\title{Cross-Polarization Effects in Sheared 2D Grating Couplers in a Photonic BiCMOS Technology}
\author{Galina Georgieva$^{1}$\thanks{E-mail: galina.georgieva@tu-berlin.de}, Karsten Voigt$^{1}$, Christian Mai$^{2}$, Pascal M. Seiler$^{1}$,  Klaus Petermann$^{1}$ and Lars Zimmermann$^{1,2}$}
\inst{$^{1}$Technische Universität Berlin, Office HFT4, Einsteinufer 25, 10587 Berlin, Germany\\
$^{2}$IHP - Leibnitz Institut für innovative Mikroelektronik, Im Technologiepark 25, 15236 Frankfurt (Oder)}
\abst{We investigate numerically and experimentally sheared 2D grating couplers in a photonic BiCMOS technology with a focus on their splitting behavior. Two realization forms of a waveguide-to-grating shear angle are considered. The cross-polarization used as a figure-of-merit is shown to be strongly dependent on the grating perturbation strength and is a crucial limitation not only for the grating splitting performance, but also for its coupling efficiency.}

\begin{document}
\maketitle

\section{Introduction}
Intra data center optical communication systems rely exclusively on low-cost technologies using a direct detection scheme. However, the increasing demand on even higher data rates poses the question on the adoption of coherent systems based e.g. on single- or dual-polarization quadrature phase shift keying (QPSK) or more generally on quadrature amplitude modulation (QAM). Such modulation formats feature the advantage of a better spectral efficiency \cite{Pito}. On the other hand, coherent systems in data centers will require a technology, which is able to reduce their total cost to a reasonable level. For that reason, silicon photonics keeps receiving a strong interest from both industry and research, as it makes use of existing CMOS or BiCMOS foundries and promises for a cost-effective co-integration of electronics and photonics \cite{AMek,LZ}. Monolithically integrated QPSK receivers have already been demonstrated for both single- and dual-polarization operation \cite{GWin,Doerr}.

In dual-polarization (DP) QPSK or QAM receivers, the utilization of two-dimensional grating couplers (2D GC) is of particular interest, as they enable simultaneous coupling and polarization splitting of an optical signal with two polarization states. However, the first attempts for their realization were lacking a good coupling efficiency, reaching \unit[20]{\%} - \unit[25]{\%} corresponding to an insertion loss  between \unit[-7]{dB} and \unit[-6]{dB} \cite{DTail,WBog,VLaere}. Since then, many attempts have been made to reduce the coupling loss. 2D GCs with \unit[-3.2]{dB} were demonstrated \cite{Boeuf}, which could be further improved up to around \unit[-2]{dB} by the adoption of a double-SOI substrate acting as a Bragg mirror \cite{LVers} or - of a backside metal mirror \cite{Luo}. Without bottom reflection mirrors, 2D GCs in \unit[220]{nm} silicon-on-insulator (SOI) technology typically show an insertion loss between \unit[-5]{dB} and \unit[-4]{dB} \cite{JZou,Leti}.

Another thoroughly examined issue in 2D GCs is the polarization-dependent loss (PDL). A possibility for its minimization was proposed by the adoption of a phase shifter \cite{Halir}. Luxtera introduced first a modified `clover' scatterer shape \cite{AMekGC}, which inspired multiple investigations on the impact of the scatterer shape on the PDL in both C- and O-bands \cite{SPlan, Zou, Sobu1}. Also a 2D GC with fabrication tolerant stretched cylinders was shown \cite{Xue}. The PDLs reported are as low as 0.2 - \unit[0.3]{dB}. 

While the 2D GC coupling efficiency and PDL have been a subject of a strong interest, effects concerning the 2D GC splitting performance remained eclipsed. In early publications \cite{DTail, WBog} the 2D GC splitting behavior was reported in terms of crosstalk. The values reached were below \unit[-15]{dB}. However, the crosstalk was not examined on any parameter dependence. In a more recent publication, the impact of the scatterer shape was discussed \cite{Sobu2}. Our work here aims to conduct a thorough analysis on the splitting performance of 2D GCs, taking into account multiple dependencies. Our figure-of-merit - the cross-polarization (short: cross-pol.) is defined as the portion of power of one of the polarizations (e.g. the $y$-polarization), which is converted to the other one (e.g. the $x$-polarization) by the 2D GC. Hereby, it is of particular interest (a) how much of the cross-pol. power is coupled from a single-mode fiber (SMF) into the 2D GC and vice versa and (b) how much of the 2D GC power loss is attributed to the cross-pol.

The examined 2D GCs are here referred to as sheared 2D GCs. In the literature, a waveguide-to-grating shear angle has been often reported in 2D GCs \cite{VLaere, Zou, Sobu1, Sobu2, Zou2}. Previously, we described a systematic design procedure of 2D GCs with a shear angle \cite{GC_GG}, which is used for the designs considered here. We distinguish two types of sheared 2D GCs, in dependence on how the shear angle is practically realized. Previously, we investigated numerically and experimentally the cross-pol. in one of the two types in its dependence on two parameters - the shear angle and the etch depth, which is a measure for the grating perturbation strength \cite{MOC_GG}. Our work is here extended with the same analyses on the other type of sheared 2D GCs. In addition, we provide a more comprehensive explanation of the basic theory and the simulation results as well as of the principle of our experiment. We further discuss the possible impact of the cross-pol. on system level. 

In Sec. \ref{sec:methods} we first briefly describe the design of BiCMOS integrated sheared 2D GGs and the working principle of the device fabricated for the experimental evaluation of the cross-pol. In the following Sec. \ref{sec:results}, numerical results outline our expectations on the sheared 2D GCs splitting behavior, depending on their type, shear angle and etch depth. Next, statistical experimental results are used to confirm our observations. The test samples used here are fabricated in a full photonic BiCMOS flow based on \unit[220]{nm} SOI, using a \unit[248]{nm} deep UV lithography and comprising the complete backend of line stack. In addition, we analyze the potential impact of different cross-pol. levels on a system using DP QAM modulation formats. The results finally summarized in Sec. \ref{sec:conclusions} point out the importance of a deep understanding of the cross-pol. effects for both 2D GC design and low bit-error-rate (BER) coherent transmission.

\section{Methods}
\label{sec:methods}

In this section, we describe the specifics of the platform, which the proposed 2D GCs are intended for. Afterwards, the numerical design procedure is outlined, followed by an explanation of our experimental approach for the cross-pol. characterization.

\subsection{Integration Platform}
Many research groups investigating GCs work with optimized buried oxide or top cladding layers. The results achieved in such cases, however, are not automatically applicable to a (Bi)CMOS foundry. In addition, experimental evidence is often based on measurements of a single device, which raises questions regarding repeatability. The devices shown in our paper have the advantage of being fully integrated in a photonic BiCMOS \cite{Knoll} fabrication flow. They meet the requirements of the used platform. Moreover, we are able to perform a statistical analysis of their behavior and discuss the large-scale repeatability of the obtained results. 

The photonic BiCMOS platform \cite{Knoll} used here integrates monolithically bulk-Si electronics in parallel with SOI photonics in the frontend of line (FEOL). Both electronic and photonic components are covered by a backend of line (BEOL) stack of about \SI{13}{\micro \metre} height, which mainly comprises multiple SiO$_2$ layers of different thickness and optical properties. In addition, the platform offers five metal layers, which can be used to realize the shortest interconnects between photonics and electronics. On the electronic side, very fast SiGe:C heterojunction bipolar transistors (HBT) can be designed for the accomplishment of high-speed electronic-photonic systems. 
The fixed BEOL top stack is of huge importance for a GC design and is here taken into account in all simulations. Another aspect is the fixed minimal feature size, determined by the \unit[248]{nm} deep UV lithography used. The latter defines the optimization limits of the GC scatterers size and shape.   

\subsection{Numerical Design of Sheared 2D GCs}
A 2D GC must be designed in such a way, that two orthogonal polarizations from two silicon waveguides are in- or out-coupled into the two polarizations of a SMF mode by diffraction. 
All sheared 2D GC designs here are intended for a silicon rib waveguide with \unit[220]{nm} Si on \SI{2}{\micro \metre} SiO$_2$, where the rib etch depth is the same as the grating etch depth. The simulations are performed in the C-band. The waveguide and the grating need to be integrated in the previously described photonic BiCMOS and covered by the full BEOL stack. 
\begin{figure}[H]
\centering
\includegraphics[width=\textwidth]{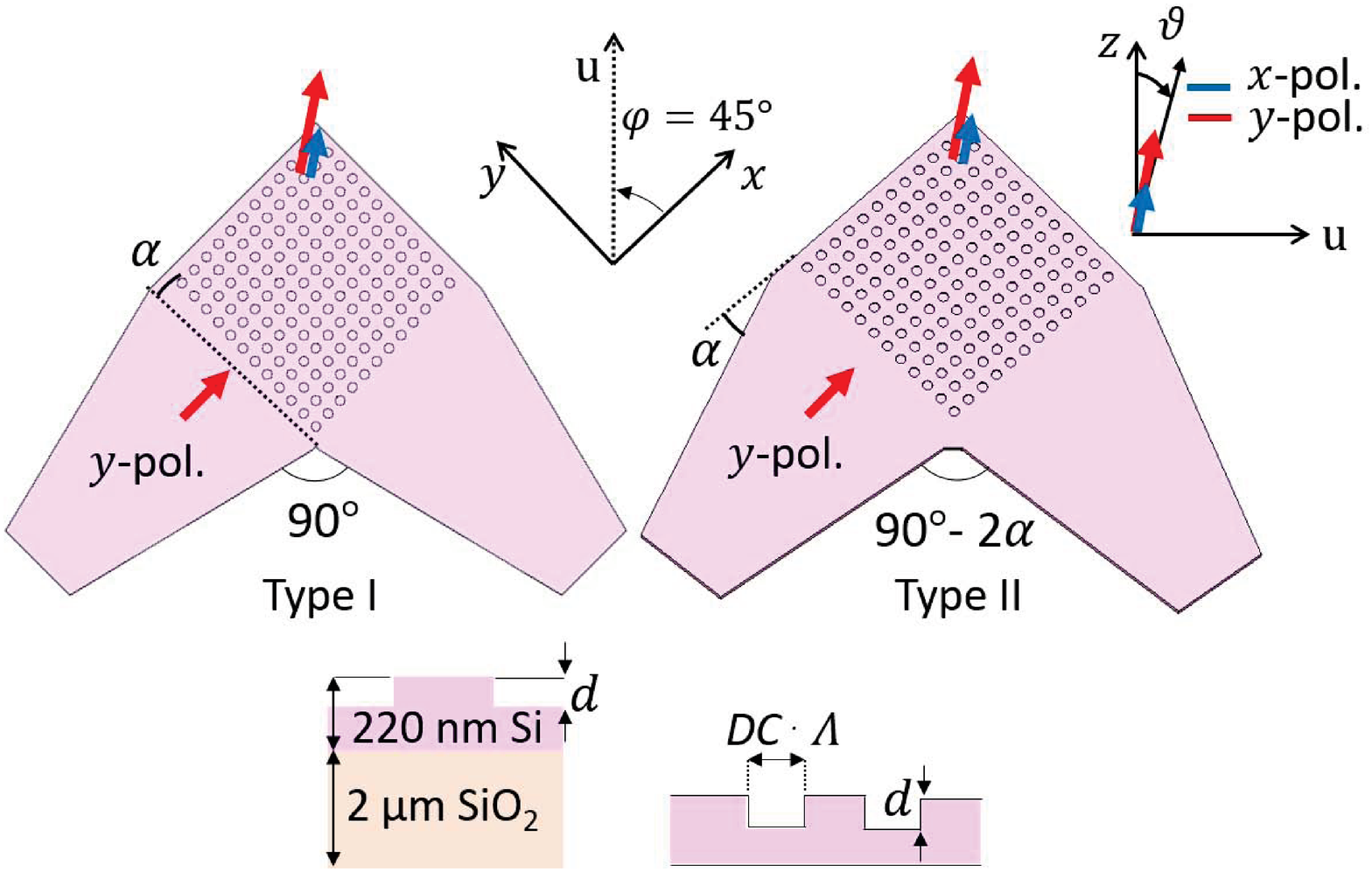}
\caption{Schematic representation of the two types of sheared 2D GCs: Type I - perpendicular waveguides and rhombus grating area, Type II - angled waveguides and square grating area. The designs are intended for a rib waveguide type. The gratings are characterized by a shear angle $\alpha$; a period $\Lambda$; a hole diameter, defined by the duty cycle $DC$ as a fraction of $\Lambda$; an etch depth $d$. The gratings diffract an input waveguide mode (here the y-pol.) under an angle $\vartheta$ with respect to the $z$-axis and along the $u$-axis, corresponding to a rotation under $\varphi = 45^\circ$. The 2D GCs possibly convert and diffract a portion of the input mode power into the $x$-pol., assigned as a cross-pol.}
\label{fig:gcs_wgs1}
\end{figure}
A waveguide schematic is shown in Fig. \ref{fig:gcs_wgs1} together with the two types of sheared 2D GC design, depending on how the waveguide-to-grating shear angle is realized. Type I gratings have perpendicular waveguides and rhombus shaped grating area, while Type II have angled waveguides and square shaped grating. The existence of a shear angle is for the reason of improving the mode field overlap between the polarizations of the two waveguides and the SMF mode. A shear angle introduces a second degree of freedom along with the grating period, which allows for a change of the mode propagation direction in both $\varphi$ and $\vartheta$ direction (see the coordinate system in Fig. \ref{fig:gcs_wgs1}). This is necessary, as a SMF must be placed at the symmetry plane $\varphi = 45^\circ$ between the two waveguides.
In our previous work \cite{GC_GG}, we described the working principle of sheared gratings and derived a 2D diffraction condition, which is used here for all designs. The condition states, that certain combinations $(\alpha, \Lambda)$ correspond to certain radiation patterns $(\varphi = 45^\circ, \vartheta)$. We showed further that the choice of larger shear angles $\alpha$ leads to a larger angle $\vartheta$ in the symmetry plane for a desired wavelength.

The coupling angles and the determination of the grating parameters depend on the grating effective refractive index $n_{\text{eff}}$. Because of the complexity of integrated gratings, which have a specific covering pattern, $n_{\text{eff}}$ is determined out of the simulated field components, by applying a 2D spatial Fourier transform. For a grating with optimized etch depth and scatterer size, corresponding to a certain $n_{\text{eff}}$, the 2D Fourier transform delivers the actual radiation angles $(\varphi, \vartheta)$ and $n_{\text{eff}}$ can be calculated from the diffraction condition. If the angles require an adaption, the grating period can be afterwards adjusted without a further change of the scatterers dimensions and $n_{\text{eff}}$.

For the numerical analysis of the cross-pol., we use a commercial time-domain solver by Simulia CST, which implements the finite-integration-technique \cite{Weiland}. We consider an out-coupling 2D GC with only one waveguide polarization excited, here the $y$-polarization (short: $y$-pol.). The radiated fields are evaluated at a tilted plane $(\varphi = 45^\circ, \vartheta)$, according to the coordinates in Fig. \ref{fig:gcs_wgs1}. We assume a standard SMF and calculate the power, which is properly coupled into its $y$-pol. The portion of power coupled into the $x$-polarization (short: $x$-pol.) is then referred to as a cross-pol. As the in- and out-coupling is reciprocal, the difference between the calculated $y$- and $x$-pol. powers in dB at a wavelength of interest corresponds to the 2D GC split ratio. A large split ratio indicates that the 2D GC is a good \emph{splitter}.  Because in practical measurements it is difficult to determine exactly the maximal transmission wavelength, we  evaluate the mean split ratio in a wavelength range of \unit[15]{nm} around the $y$-pol. maximal transmission. In this interval, the transmission changes by no more than \unit[0.3]{dB}, which is typically the accuracy limit in transmission measurements.

Furthermore, we are interested also in the question whether the 2D GC is an efficient \emph{coupler}, and consider for a wavelength of interest the power that remains inside the 2D GC without being radiated. We again calculate how much of this power loss is caused by the cross-pol.

Exact absolute values of the split ratio are challenging to simulate, because they depend on the numerical accuracy. Generally, a large amount of grid cells per wavelength is required for results with high accuracy, which is however connected to an extremely long simulation time. Nevertheless, all results are calculated with the same accuracy (15 grid cells per wavelength) and allow for comparison of the structures with each other. 
It is important to note that 2D GCs of Type I and Type II require a different kind of mode excitation. While the waveguide mode in Type I can be calculated by a mode port, Type II has angled waveguides and an angled port is not supported by the time-domain solver. In that case, a waveguide mode is calculated externally and afterwards imported and tilted under the desired shear angle. Minor, differences in the numeric results between the two types of 2D GCs may thus result because of the different simulation procedure.

\subsection{Device for a Cross-Polarization Measurement}
\begin{figure}[H]
\centering
\includegraphics[width = \textwidth]{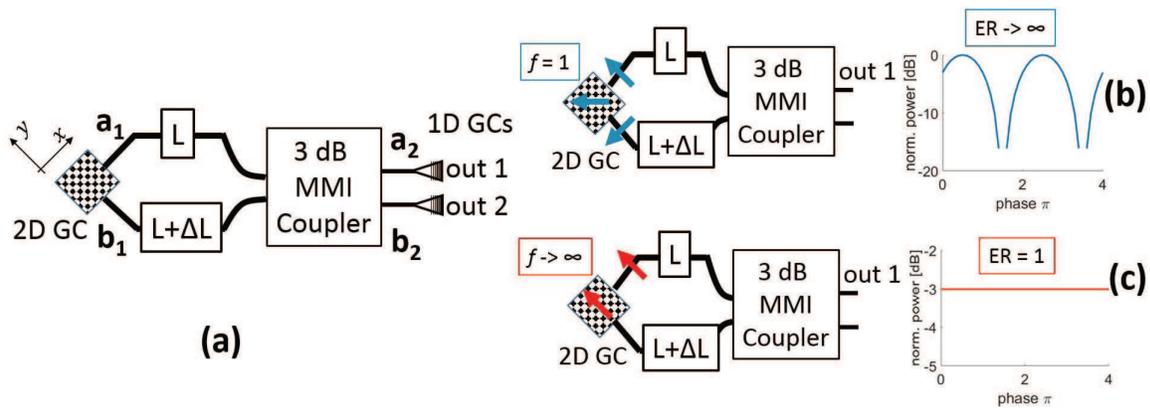}
\caption{(a) Schematic of a device for cross-pol. measurement, (b) Incident polarization for a split ratio $f = 1$, leading to an ER of infinity, (c) Incident polarization for a split ratio $f \rightarrow \infty$, leading to an ER of 1 and vanishing resonances.}
\label{fig:device}
\end{figure}

Fig. \ref{fig:device} (a) shows a schematic of the structure used for the 2D GC cross-pol. characterization. It comprises a 2D GC under study, \SI{500}{\micro \metre} long linear tapers for adiabatic mode conversion to \unit[500]{nm} wide waveguides (not depicted), a low-loss delay line of \SI{200}{\micro \metre} length on the lower arm, a multi-mode interference (MMI) coupler and two focusing 1D GCs at the output. The MMI coupler split ratio is wavelength dependent, but the variation is small.

The device working principle is the following - the 2D GC splits an incident wave of an arbitrary polarization into the two waveguide arms. If we denote these amplitudes as $\mathbf{a_1}$ and $\mathbf{b_1}$, the ratio of their squares corresponds to the grating split ratio $f$, i.e to the power ratio between the polarizations in the two waveguides. We further normalize the amplitudes to the 2D GC input power 
\begin{equation*}
f = \frac{\mathbf{|a_1|}^2}{\mathbf{|b_1|}^2}, \qquad \mathbf{|a_1|}^2 + \mathbf{|b_1|}^2 = 1.
\end{equation*}
We choose such coordinates that the split ratio here is defined as the $y$- to $x$-pol. power ratio similar to the simulated case. After the delay line and the MMI, the two signal paths interfere constructively or destructively at the outputs, depending on the wavelength and the corresponding phase delay $\Delta \phi$ in the delay line. The ratio between the maximal and minimal signal transmission is known as an extinction ratio (ER) and can be measured after out-coupling with a 1D GC. 

From the ER we are able to estimate the 2D GC split ratio by using a simple matrix model \cite{KVoigt}. The relation between the input waves $\mathbf{a_1}$ and $\mathbf{b_1}$ and the output waves $\mathbf{a_2}$ and $\mathbf{b_2}$ can be mathematically written as a multiplication of a matrix $\mathbf{T_{MMI}}$ describing the MMI coupler and a matrix $\mathbf{T}_{\Delta \phi}$ describing the delay line:

\begin{align}
\begin{bmatrix} \mathbf{a_2} \\ \mathbf{b_2} \end{bmatrix} &= \mathbf{T_{MMI}}\cdot \mathbf{T}_{\Delta \phi} \begin{bmatrix} \mathbf{a_1} \\ \mathbf{b_1} \end{bmatrix} \quad \text{with}\\
\mathbf{T_{MMI}} &= \frac{1}{\sqrt{2}} \begin{bmatrix} 1 & j \\ j & 1 \end{bmatrix}, \quad%
\mathbf{T}_{\Delta \phi} =  \begin{bmatrix} \mathrm{e}^{j\Delta \phi} & 0 \\ 0 & 1 \end{bmatrix}.
\end{align}

If we take e.g. the first output, the power measured there is proportional to the square of the wave amplitude $\mathbf{a_2}$
\begin{equation}
P_{\text{out}} \sim |\mathbf{a_2}|^2 = \frac{1}{2}\Bigl( \mathbf{|a_1|}^2 + \mathbf{|b_1|}^2 + 2 \mathbf{|a_1|} \mathbf{|b_1|} \sin \Delta \phi \Bigr).
\end{equation}
The maximum and the minimum of $P_{\text{out}}$ can be obtained by maximizing or minimizing $\sin \Delta \phi$ and their ratio results in the ER. If we further replace $\mathbf{a_1}$ and $\mathbf{b_1}$ through their split ratio $f$, we obtain 
\begin{equation}
\mathrm{ER} = \Biggl( \frac{1+ \sqrt{f}}{1 - \sqrt{f}}  \Biggr)^2.
\label{eq:ER}
\end{equation}

Measured ERs can be easily translated into split ratios. It is still an open question, how we determine if the 2D GC splits the polarizations well. The answer is illustrated by two examples, which do not consider the GCs filter spectra. In Fig. \ref{fig:device} (b), the incident polarization is such that the signal is equally split between the two waveguides. The split ratio is therefore 1 and from (\ref{eq:ER}) we obtain an ER of infinity. The ER is, however, limited in real systems, so we consider another case in Fig. \ref{fig:device} (c). This time, the incident polarization is such that all the power is coupled in one of the waveguides. The example is chosen to match the simulated case. The split ratio is now $f \rightarrow \infty$ and the ER is 1, which means that minimal and maximal levels are the same and no resonances result. In our experiment, we look therefore for a polarization, which leads to a minimal ER. Ideally, smooth transmission without resonances should result. Out of the minimal ER, we can estimate the best achievable split ratio of the considered structure.  

The principle of our measurement may first look rather complicated, but as we perform manual measurements, it has two obvious advantages. First, the signals at the two MMI coupler outputs are the same and only phase shifted. Thus, it is sufficient to measure only one of the outputs, which reduces the time for the measurement and/or the setup complexity. Second, if we measure on two outputs, we need to guarantee that they are equally well coupled, which is practically very difficult in manual measurements. In this experiment, the in- and out-coupling efficiency has no impact on the determined split ratios. Also the mechanical coupling stability during the long wavelength sweep (typically 3-4 minutes) is less crucial.

\subsection{Experimental Approach and Setup}
The experimental verification of the numerically predicted behavior is carried out by manual wafer measurements. We compare structures of Type I and II as well structures with different shear angles $\alpha = 2^\circ$ and $\alpha = 3^\circ$ on the same wafer. For the comparison of the structures with different etch depths $d = \unit[90]{nm}$ and $d = \unit[120]{nm}$, two separate wafers are available. Fabrication variations on these wafers may be different, so we need to compare in the end an averaged split ratio over a certain number of chips.  We use a laser source Agilent 81960 with a wavelength range from \unit[1505]{nm} to \unit[1625]{nm}, followed by a programmable polarization controller Agilent 8169A and two standard SMFs for the in- and out-coupling together with their alignment equipment. The signal is finally detected by a power sensor Agilent 81634B. Back-to-back reference structures are used for the determination of the proper coupling angles at both sides. They are chosen such, that the 2D GC and the 1D GC have a maximal coupling centered at the same wavelength. For the MMI measurement, we scan the polarization angles on the Poincaré sphere and determine the polarization, for which the minimal ER results. The wavelength sweep is in \unit[1]{pm} increment and 10 chips are considered for the statistics. Similar to the simulations, the split ratio is averaged in a wavelength range of \unit[15]{nm} around the maximal transmission and afterwards averaged over the 10 chips. Note that the first decimal place of the split ratio varies depending on how the averaging is performed - by first averaging the ERs or by converting them into split rations and then taking the mean value. Here, we first calculate the split ratios out of the ERs for all chips and average them in the end.

\section{Results and Discussion}
\label{sec:results}

In this section, numerical and experimental results are used to discuss the impact of different design parameters on the cross-pol. effects in sheared 2D GCs. The 2D GC tilt angle is fixed to  $\vartheta = 8^\circ$, so we can compare different devices only around different central wavelengths. The constant angle is necessary to ensure similar coupling conditions in all measurements. In the last subsection, we move from device to a system aspect investigation of the cross-pol. effects by taking them into account in DP QAM transmission simulations.

\subsection{Numerical Results}
\begin{figure}[H]
\centering
\includegraphics[width = .8\textwidth]{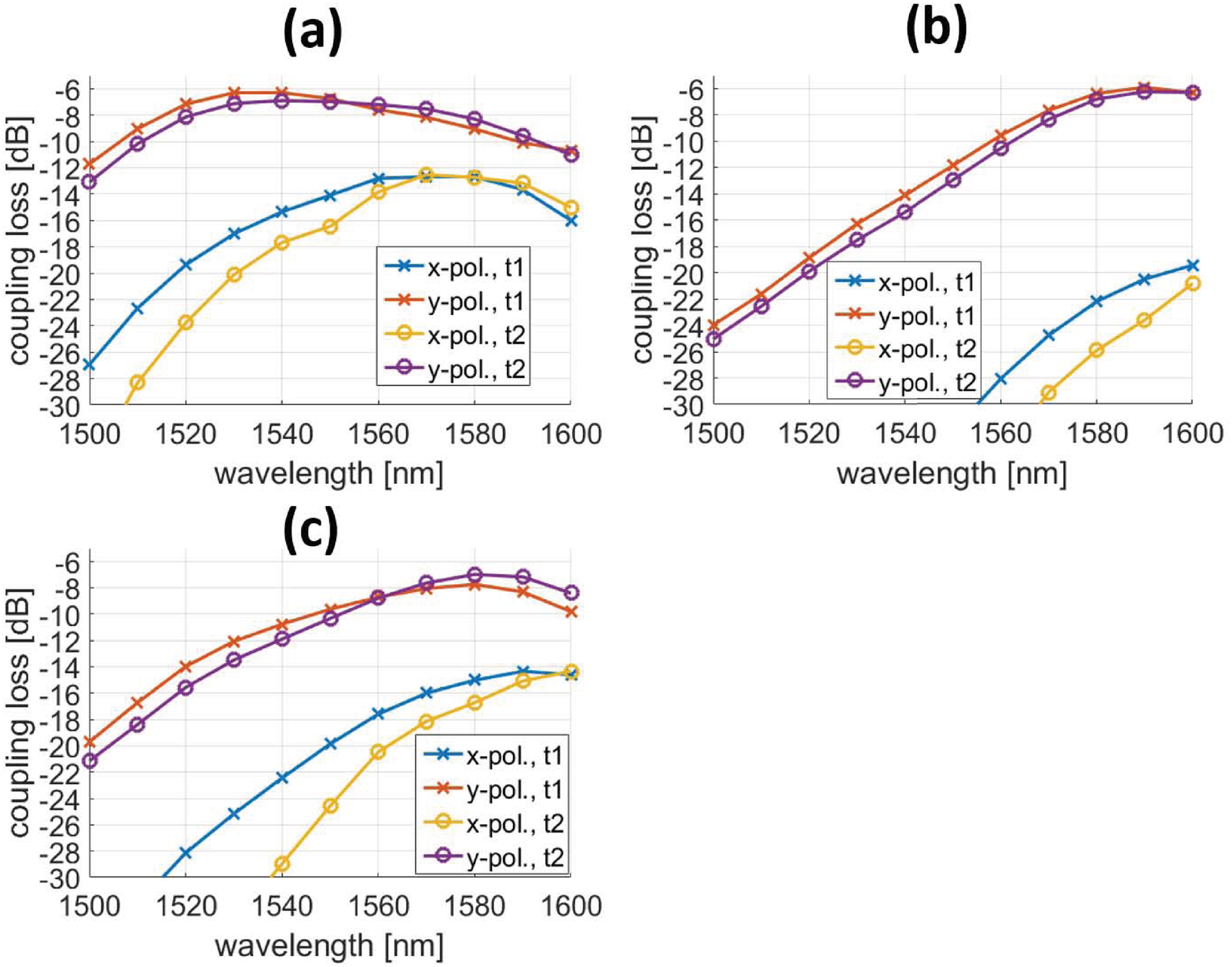}
\caption{Simulated out-coupling spectra at $\varphi = 45^\circ, \vartheta = 8^\circ$ of sheared 2D GC for the comparison of the split ratio of three cases for two types of gratings - t1 (Type I with a rhombus-shaped grating area) and t2 (Type II with angled waveguides). (a) Shear angle $\alpha = 2^\circ$, etch depth $d = \unit[120]{nm}$ (b) Shear angle $\alpha = 2^\circ$, etch depth $d = \unit[90]{nm}$ (c) Shear angle $\alpha = 3^\circ$, etch depth $d = \unit[120]{nm}$.}
\label{fig:sim_all2dgc_1}
\end{figure}

In all simulations, we use the waveguide with $y$-pol. as an input and optimize the grating parameters for its out-coupling in direction ($\varphi = 45^\circ, \vartheta$), where the resulting $\vartheta$ depends on the choice of a shear angle $\alpha$ and a grating period $\Lambda$. The duty cycle $DC$ is in any case 0.71. For $\alpha = 2^\circ$ and $\Lambda = \unit[622]{nm}$ and an etch depth $d = \unit[90/120]{nm}$, we obtain for both grating types $\varphi = 45^\circ, \vartheta \approx 8^\circ$ at a wavelength of 1590/\unit[1550]{nm}. For  $d = \unit[120]{nm}, \alpha = 3^\circ$ and $\Lambda = \unit[636]{nm}$, the tilt angle increases to $\vartheta \approx 12^\circ$ at \unit[1550]{nm}.  For a fixed angle $\vartheta = 8^\circ$, the central wavelength will be shifted to \unit[1580]{nm}.  

Fig. \ref{fig:sim_all2dgc_1} shows the simulated out-coupling spectra for the two types of sheared gratings, comparing different cases. In any of them, we see that the two types of gratings Type I and Type II (short for the figures - t1 and t2) differ only slightly from each other and the differences can be caused by the different simulation procedures mentioned previously. Another common characteristic in all cases is that the cross-pol. ($x$-pol.) central wavelength is shifted to larger wavelengths compared to the $y$-pol. The shift is larger in the case of the gratings of Type II. Fig. \ref{fig:sim_all2dgc_1} (a) and (b) compare the geometries with different etch depths, the remaining parameters fixed. Fig. \ref{fig:sim_all2dgc_1} (a) and (c) compare models comprising different shear angles. For each structure, a mean split ratio in a \unit[15]{nm} interval around the maximal transmission wavelength is given in Tab. \ref{tab:2DGCsplit}. Because of the larger cross-pol. wavelength shift for the Type II grating, its split ratios appear better than for the Type I grating. Comparing the structures with shear angles $2^\circ$ and $3^\circ$, we see that the latter shows around 1-\unit[2]{dB} worse split ratio. This shear angle requires, however, a different grating period, so the worse behavior may be caused not by the shear angle alone, but by the combination of e.g. a certain shear angle with a certain grating period. 

\begin{table}[H]
\caption{Numerically estimated split ratios $f$ in a \unit[15]{nm} wavelength range near the maximal $y$-pol. transmission for different out-coupling sheared 2D GCs. The coupling angle is $8^\circ$.}
\label{tab:2DGCsplit}
\begin{tabular}{c|c|c|c|c}
\Hline % To generate a thicker line than \hline
Shear angle $\alpha$  & Etch depth $d$ [nm]  & Wavelength [nm] &2D GC Type & Split ratio  $f$ [dB]  \\ \hline
2$^\circ$ & 120 & 1530-1545 & I & 9.5 \\
		  &		&      & II & 11.5 \\
   
\hline
3$^\circ$ & 120 & 1570-1585 & I & 7.4  \\
		  &		&      & II & 9.8 \\
\hline
2$^\circ$ & 90 & 1585-1600 & I & 14.2  \\
		  &		&      & II & 16.5 \\
\Hline
\end{tabular}
\end{table}

\begin{figure}[H]
\centering
\includegraphics[width = .95\textwidth]{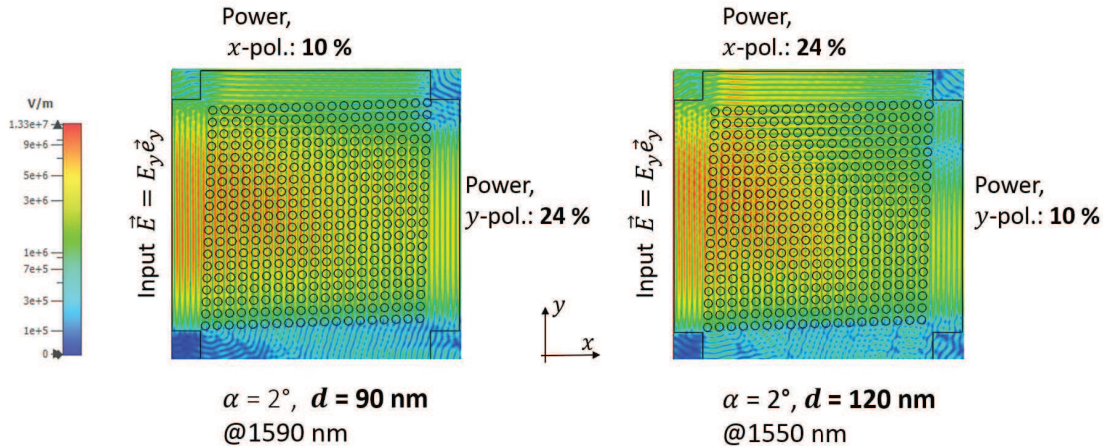}
\caption{Normalized power remaining inside a 2D GC with a shear angle $\alpha = 2^\circ$. The input mode is purely $y$-polarized. After propagating through the 2D GC, a portion of its power is radiated under $\varphi = 45^\circ, \vartheta = 8^\circ$ and coupled into a SMF. The power loss is caused by a radiation into the substrate (not represented) and remaining power that propagates further inside the 2D GC. This propagating mode has two polarization states. For an etch depth $d =\unit[90]{nm}$, the power remains mainly within the initial $y$-pol. (\unit[24]{\%}) and \unit[10]{\%} are converted into the cross-pol.. For an etch depth $d =\unit[120]{nm}$ the power loss is mainly caused by the cross-pol. conversion (\unit[24]{\%}).}
\label{fig:2DGC_lossIn} 
\end{figure}

A comparison between the structures with different etch depths in Tab. \ref{tab:2DGCsplit} shows clearly that a larger etch depth, corresponding to a stronger grating perturbation, leads to about 4-\unit[5]{dB} worse split ratio. For a good splitting functionality, shallowly etched 2D GCs should be preferred. However, other than the shear angle, the etch depth has a strong impact on the grating out-coupled power. If the perturbation strength is too weak, a large amount of power remains guided and is not radiated by the grating. We consider this kind of loss and compare how much of the normalized input power remains inside the structure for the two different etch depths. 
Consider the numerical example in Fig. \ref{fig:2DGC_lossIn}. On the left hand side, we have an input mode, which is purely $y$-pol. While this mode is propagating through the 2D GC, a portion of its power is coupled out towards a SMF. The geometries are such that the out-coupling angles are $\varphi = 45^\circ, \vartheta = 8^\circ$ for a wavelength of \unit[1590]{nm} and an etch depth of \unit[90]{nm} or for a wavelength of \unit[1550]{nm} and an etch depth of \unit[120]{nm}. Power losses are caused by radiation towards the substrate (not represented) and by the power guided in the structure. In Fig. \ref{fig:2DGC_lossIn}, we clearly see that in both cases the further propagating mode has an $y$-pol. part (in the waveguide on the right hand side) and an $x$-pol. part (in the upper waveguide). For the etch depth $d =\unit[90]{nm}$ the power remains mainly within the initial $y$-pol. (\unit[24]{\%}) and \unit[10]{\%} are converted into the cross-pol. On the other hand, for an etch depth $d =\unit[120]{nm}$ the power loss is mainly caused by the cross-pol. conversion (\unit[24]{\%}) and overall we do not increase the amount of the radiated power. The cross-pol. conversion increases with increasing grating perturbation strength, which is compared to 1D GCs an additional loss mechanism limiting the maximal achievable coupling efficiency.

Finally, we have a look over the split ratio variation, if the fabricated grating holes show an asymmetry and have elliptical instead of circular shape. This would lead to a different wavelength shift of the spectra of the $y$-pol. and the cross-pol. If the holes are stretched or shrunk by \unit[20]{nm} in $y$-direction we have around $\mp \unit[1]{dB}$ variation of the split ratio. The variation is small compared to the expected split ratio difference for different etch depths, but of similar order with the expected split ratio difference between 2D GCs of different types or shear angles. 
\subsection{Experimental Results}
\begin{figure}[H]
\centering
\includegraphics[width = .8\textwidth]{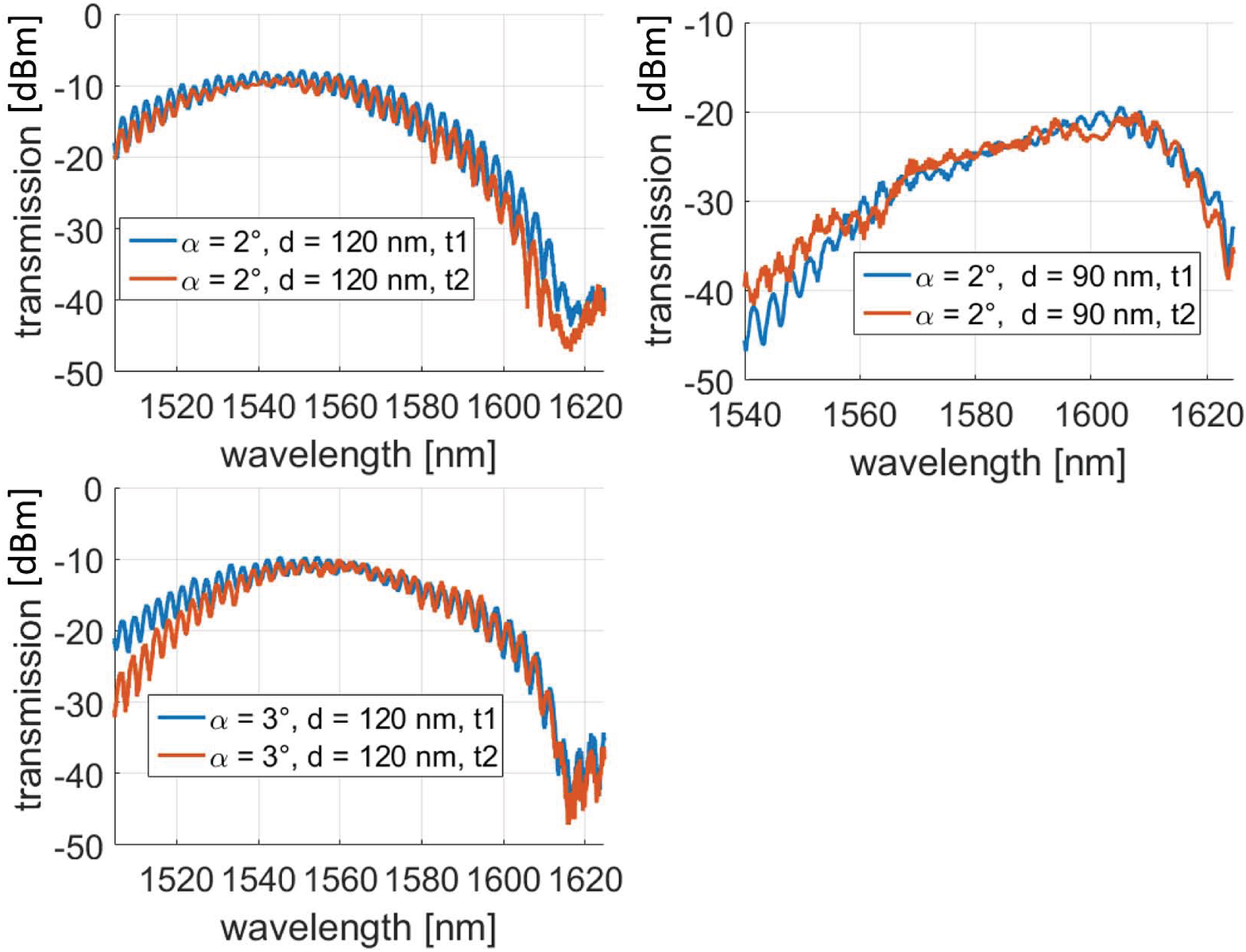}
\caption{Exemplary measured interferometric curves of devices with sheared 2D GCs used for the comparison of the split ratio of three cases for two types of gratings - t1 (Type I with rhombus-shaped grating area) and t2 (Type II with angled waveguides).}
\label{fig:meas_all2dgc_1}
\end{figure}

The structures discussed in the previous subsection are now experimentally investigated. Fig. \ref{fig:meas_all2dgc_1} shows on example interferometric curves for each of the devices considered. The polarization is chosen such, that a minimal ER results. The transmission includes the coupling losses at the input and the output. Compared to our previous results \cite{MOC_GG} we managed to further minimize the ER for each of the structures, by scanning more precisely the Poincaré sphere. This does not change our general observations. Another difference here is the wider wavelength range of \unit[15]{nm} around the maximal transmission. We also repeated our previous measurement on the Type I 2D GCs, in order to make the two types of 2D GCs comparable to each other. 
It is important to note that split ratios below \unit[20]{dB}, which correspond to ERs of below \unit[1.5]{dB} become very difficult to measure, because other effects like Fabry-Perot resonances and noise start to predominate and make the determination of the ER difficult.

\begin{table}[H]
\caption{Experimentally estimated mean split ratios $\overline{f}$ in a \unit[15]{nm} wavelength range near the maximal $y$-pol. transmission for different out-coupling sheared 2D GCs. The values are averaged over 10 chips with a standard deviation $\sigma$. The coupling angle is $8^\circ$.}
\label{tab:2DGCsplit_meas}
\begin{tabular}{c|c|c|c|c}
\Hline % To generate a thicker line than \hline
Shear angle $\alpha$  & Etch depth $d$ [nm]  & Wavelength [nm] &2D GC Type &  $\overline{f} \pm  \sigma$ [dB]  \\ \hline
2$^\circ$ & 120 & 1527-1542 & I & $16.4 \pm 1.4$ \\
		  &		&      & II & $17.7 \pm 1.6$ \\
   
\hline
3$^\circ$ & 120 & 1542-1557 & I & $15.3 \pm 1.0 $  \\
		  &		&      & II & $17.7 \pm 0.9$ \\
\hline
2$^\circ$ & 90 & 1579-1594 & I & $19.7 \pm 1.1$  \\
		  &		&      & II & $19.1 \pm 1.7$ \\
\Hline
\end{tabular}
\end{table}

In Tab. \ref{tab:2DGCsplit_meas} the averaged split ratios in a \unit[15]{nm} wavelength range for each of the cases are summarized. The mean split ratio results from the averaging over 10 chips and the standard deviation $\sigma$ is given as well. All structures, which are directly compared to each other, show a similar standard deviation. The differences between structures of Type I and Type II vary from one structure to another and we cannot generally state that one of the types has an advantageous splitting behavior. The lack of systematic difference can be caused by variations of the hole shapes, causing various 2D GC asymmetries. Such variations can make differences between the gratings with the different shear angles $\alpha = 2^\circ, 3^\circ$ also difficult to be observed. Being difficult to distinguish from other fabrication variations, the differences between the 2D GCs of different types or with different shear angle appear to be of minor importance.  

The split ratio for a certain etch depth appears to be of similar order, independent of the particular 2D GC geometry. Compared to the simulations, the difference between the split ratios for $d = \unit[120]{nm}$ and $d = \unit[90]{nm}$ is lower - between  \unit[1]{dB} and \unit[3]{dB} and better pronounced for the 2D GCs of Type I. The deviations from the simulation can be caused again by different asymmetry variation of the gratings. Nevertheless, the distinction between structures with different etch depths is still measurable, which shows that the grating perturbation strength has the most pronounced impact on the 2D GC splitting performance. 

If we discuss the absolute values of the split ratios, we have to point out that they depend on the channel width around a wavelength of choice. We consider only a narrow channel around a maximal transmission wavelength. The estimated split ratios obtained by the experiment reach levels that are better than predicted numerically. This can result from several reasons. The first one is the previously mentioned limited accuracy of the numerical results, because of the chosen wavelength and spatial resolution. Another reason can be the fact that in simulations the impact of the tapers is not taken into account. The desired polarization and the cross-pol. do not have generally the same coupling angles ($\varphi, \vartheta$), which is evident by their shifted coupling spectra. Thus, it is possible that the cross-pol. has a higher taper loss, especially if it enters the taper under a non-perpendicular angle. In future, the possible impact of tapers and waveguides needs to be investigated in more detail.

\subsection{DP QAM Systems with Cross-Polarization}
\begin{figure}[H]
\centering
\includegraphics[width = \textwidth]{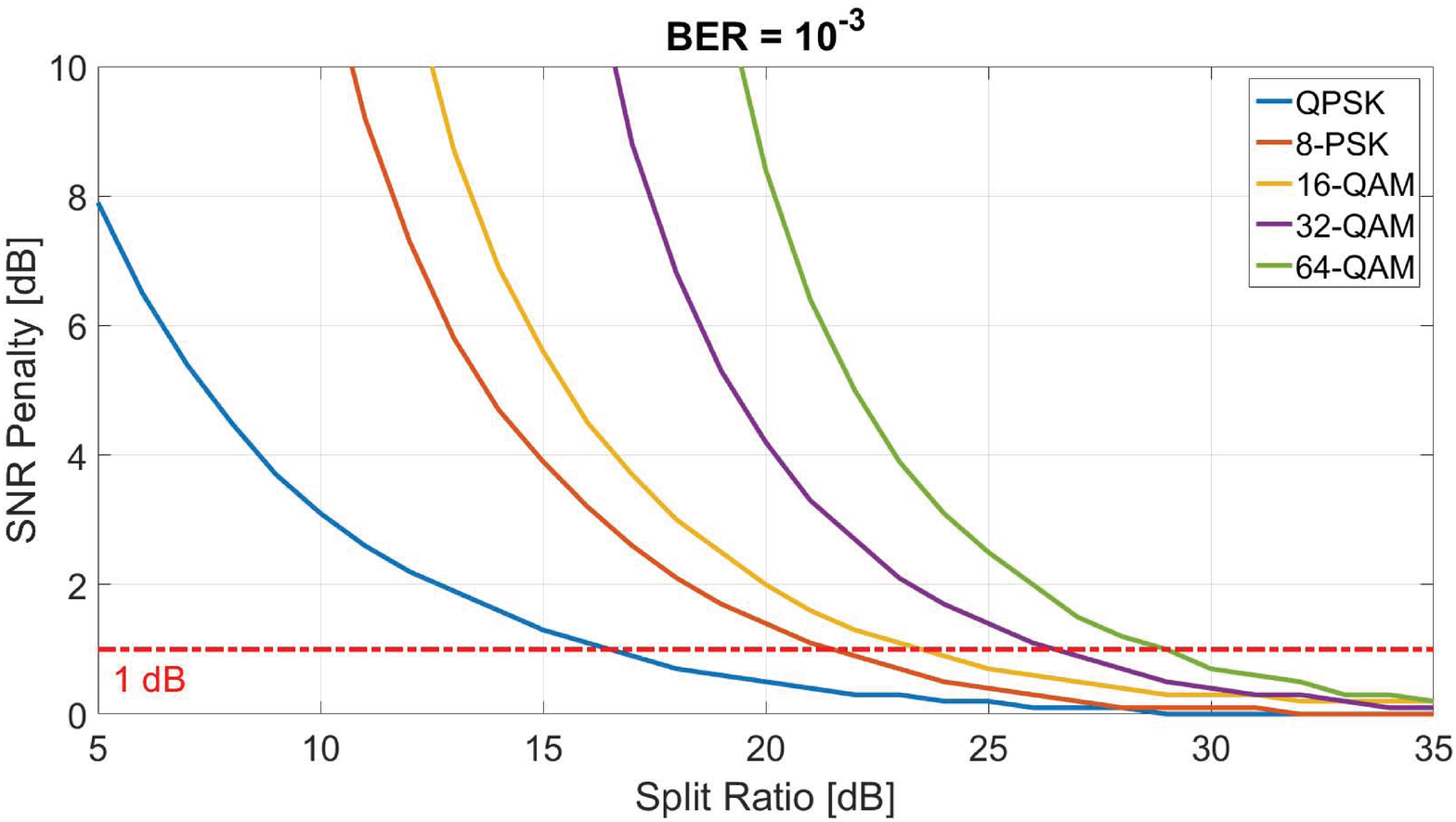}
\caption{Exemplary simulation of the SNR penalty for reaching the forward error correction (FEC) limit of BER = 10$^{-3}$, plotted over different 2D GC split ratios. The considered DP QAM formats are QPSK, 8-PSK, 16-QAM, 32-QAM and 64-QAM.}
\label{fig:snr_split}
\end{figure}

In an integrated DP coherent detection receiver, a 2D GC cross-pol. resulting in a limited polarization split ratio can be expressed as a crosstalk between two data streams with the same source. The system behavior in such a case is comparable to the in-band crosstalk, discussed in long haul transmission systems \cite{PJWinzer}. We perform an exemplary simulation, where the split ratio is varied and the SNR penalty for reaching the forward error correction (FEC) limit of BER = 10$^{-3}$ is determined. A \unit[28]{Gbaud} transmission using a laser with a linewidth of \unit[100]{kHz} is assumed (local oscillator effects are neglected). Polarization mode dispersion (PMD) is not taken into account. Fig. \ref{fig:snr_split} shows the SNR penalty over the 2D GC split ratio for different QAM formats. If we permit a \unit[1]{dB} penalty (corresponding to the red dashed line in Fig.  \ref{fig:snr_split}), we can estimate the minimal split ratio required for the desired QAM format. The results shown in this example correspond closely to the findings in the case of a fiber transmission link \cite{PJWinzer}. For QPSK, 8-PSK, 16-QAM, 32-QAM and 64-QAM, the required split ratios are \unit[16.5]{dB}, \unit[21.5]{dB}, \unit[23.5]{dB}, \unit[26.5]{dB} and \unit[29]{dB}. The 2D GC designs discussed here reach mostly split ratios better than \unit[16.5]{dB}, which make them suitable for DP QPSK. However, for higher order modulation formats the limited split ratio becomes more critical. Therefore, the suppression of cross-pol. effects in 2D GCs is crucial for the realization of integrated higher order DP QAM systems.

\section{Conclusions}
\label{sec:conclusions}

We analyzed numerically and experimentally cross-polarization effects in sheared 2D grating couplers. The focus was set on the gratings splitting behavior, but the grating coupling efficiency was shown to be affected by the cross-polarization as well.

Numerical results predicted small differences in the splitting performance of sheared 2D GCs of different types or with different shear angles. Experimentally, such differences could not be systematically measured. Their strength is therefore of the same order like effects because of fabrication deviations on the wafer. Thus, the choice of a 2D GC type or a shear angle plays a less significant role for the 2D GC splitting functionality.

A comparison between 2D GCs with a variable etch depth corresponding to a variable perturbation strength showed both numerically and experimentally that deeply etched gratings have a worse splitting performance. In addition, simulations showed that increased perturbation strength leads to a large amount of power loss, caused by a cross-polarization further propagating in the grating. The cross-polarization restricts therefore the maximal achievable out-coupled power, which is an important limitation for the 2D GC coupling efficiency.

Finally, we estimated the potential impact of different levels of cross-polarization, when a 2D GC is used as a part of a DP QAM receiver. For a QPSK modulation, the demands on the SNR penalty for a desired BER can be met by the current 2D GCs. The adoption of higher order modulation formats requires further improvement of the 2D GCs splitting functionality.

Experimentally, the splitting performance was less affected by cross-polarization effects than numerically predicted. In future, the cross-polarization propagation behavior and the influence of other components like tapers, waveguides and bends are to be studied in more detail. Furthermore, possibilities for optimizing the 2D GC perturbation strength to achieve both high efficiency and low cross-polarization are to be investigated.

\end{document}